\documentclass[]{interact}

\usepackage{epstopdf}
\usepackage[caption=false]{subfig}

\usepackage[numbers,sort&compress]{natbib}
\bibpunct[, ]{[}{]}{,}{n}{,}{,}
\renewcommand\bibfont{\fontsize{10}{12}\selectfont}

\theoremstyle{plain}

\theoremstyle{definition}

\theoremstyle{remark}

\DeclareMathOperator*{\argmax}{arg\,max}

\DeclareMathSymbol{\mh}{\mathord}{operators}{`\-}
\usepackage{amsmath}
\usepackage[colorlinks,citecolor=blue,urlcolor=blue,filecolor=blue,backref=page]{hyperref}

\usepackage{algorithm} 
\usepackage{algpseudocode}
\usepackage{caption}
\usepackage{comment}
\usepackage{booktabs}
\usepackage{amsfonts}
\usepackage{pdflscape}
\usepackage{tabularx}
\usepackage{graphicx}
\graphicspath{{./img/}}

\DeclareUnicodeCharacter{2212}{-} 
\DeclareUnicodeCharacter{2113}{+} 

\begin{document}

\articletype{ARTICLE TEMPLATE}

\title{A Data Driven Bayesian Graphical Ridge Estimator}

\author{
\name{J. Smith\textsuperscript{a}\thanks{CONTACT A.~N. Author. Email: jarodsmith706@gmail.com},
M. Arashi\textsuperscript{ab} and
A. Bekker\textsuperscript{ac}}
\affil{\textsuperscript{a} Department of Statistics, University of Pretoria, Pretoria, 0002, South Africa; 
\textsuperscript{b} Department of Statistics, Faculty of Mathematical Sciences, Ferdowsi University of Mashhad, Mashhad, Iran;
\textsuperscript{c} Centre of Excellence in Mathematical and Statistical Sciences, Johannesburg 2000, South Africa}
}

\maketitle

\begin{abstract}
Bayesian methodologies prioritising accurate associations above sparsity in Gaussian graphical model (GGM) estimation remain relatively scarce in scientific literature. It is well accepted that the $\ell_2$ penalty enjoys a smaller computational footprint in GGM estimation, whilst the $\ell_1$ penalty encourages sparsity in the estimand. The Bayesian adaptive graphical lasso prior is used as a departure point in the formulation of a computationally efficient graphical ridge-type prior for events where accurate associations are prioritised over sparse representations. A novel block Gibbs sampler for simulating precision matrices is constructed using a ridge-type penalisation. The Bayesian graphical ridge-type prior is extended to a Bayesian adaptive graphical ridge-type prior. Synthetic experiments indicate that the graphical ridge-type estimators enjoy computational efficiency, in moderate dimensions, and numerical performance, for relatively non-sparse precision matrices, when compared to their lasso counterparts. The adaptive graphical ridge-type estimator is applied to cell signaling data to infer key associations between phosphorylated proteins in human T cell signalling. All computational workloads are carried out using the baygel R package.
\end{abstract}


\section{Introduction}\label{sec:introduction}
Graphical models use graph structures to model complex relationships among a multiplicity of variables. In particular, undirected graphs encode the conditional dependence structure as symmetric relationships between pairs of variables. Gaussian graphical models (GGMs) refer to undirectd graphs characterised by non-zero off-diagonal elements of the precision matrix $\mathbf{\Omega}=\mathbf{\Sigma}^{-1}$, assuming that the data is generated by a multivariate Gaussian distribution. The GGM approach for covariance selection, \cite{dempster_1972}, is ubiquitous in statistical inference owing to its ability to map conditional independencies between variables as a probabilistic graphical network, \cite{lauritzen1996graphical}. For example, GGMs are used for architecting gene expressions \cite{yin_2011_sparse}; the interactions of symptoms in psychological disorders over time \cite{epskamp_2018_gaussian}; differential networks \cite{ali_shojaie} and more recently in modeling the connectivity structures between nodes in the brain (\cite{das_2017}, \cite{hang_2020}).\\ 
\\
From a Bayesian viewpoint, statistical inference of GGMs is typically achieved through the use of hierarchical priors for simultaneous precision matrix estimation and structure learning. The $G\mathrm{Wishart}$ is a popular prior choice for $\mathbf{\Omega}$. It places positive probability mass at zero for zero off-diagonal elements and and enjoys conjugacy for Gaussian distributed data (\cite{dawid_1993_hyper}, \cite{roverato_2002_hyper}, \cite{letac_2007_wishart}). For example, (\cite{wang_2012_efficient}, \cite{wang_2015_scaling}) make use of the $G\mathrm{Wishart}$ in conjunction with independent Bernoulli priors. The latter serve as a priori  for the the binary variables responsible for edge-inclusion indicators of the undirected graph, (\cite{wang_2012_efficient}, \cite{wang_2015_scaling}). \cite{mohammadi_2015_bayesian} investigate the use of discrete uniform and truncated Poisson priors for graph structure learning. \cite{wang_2013_class} propose a class of of shrinkage priors based on a scale mixture of uniform distributions. 

Although the $G\mh\mathrm{Wishart}$ prior has enjoyed success in many applications, its computational demand and restrictive form may prove challenging (\cite{dobra_2011_bayesian}, \cite{cheng_2012_hierarchical}). In recent years, several
alternative priors for the estimation of  $\mathbf{\Omega}$ have been proposed. \cite{wong_2003_efficient} make use of point mass priors on zero off-diagonal elements and gamma priors for the diagonal elements of $\mathbf{\Omega}$. 

Alternatively, absolute continuous priors draw direct connections to penalised estimation with the aim of inducing shrinkage to zero, or near zero, of off-diagonal elements. From a classical viewpoint, there is no shortage of literature pertaining to shrinkage estimators under an assortment of penalty functions.  The popular graphical lasso (\cite{yuan_and_lin_glasso}, \cite{friedman_2008_sparse} and the adaptive graphical lasso \cite{fan_2009_network} are among the most successful. Bayesian approaches focus on posterior mode estimation and in contrast to the $G\mh\mathrm{Wishart}$, the shrinkage priors do not place positive probability mass at zero for zero off-diagonal elements. This approach allows for improved computational scalability through the use of efficient sampling algorithms, for example block Gibbs samplers. In particular, \cite{wang_2012_bayes} and \cite{khondker_2013_bayesian} describe a Bayesian graphical lasso utilising independent exponential priors for the elements of the main diagonal and independent Laplace priors for the off-diagonal. Here, the posterior mode provides the Bayesian connection to the popular graphical lasso. A graphical horseshoe prior is proposed by \cite{li_2019_graphical}, whereas \cite{li_2019_spike_slab} developed a class of continuous spike-and-slab mixture priors for joint precision estimation. It should be noted, however, that under absolutely continuous priors, these Bayesian techniques require a heuristic treatment for structure learning. 

Bayesian ridge-type estimation of $\mathbf{\Omega}$ offers a juxtaposed approach compared to its $\ell_1$ counterparts, in that accurate associations are prioritised over sparse representations (\cite{van_2016_ridge}, \cite{bilgrau_2020_targeted}). This ideology sets the stage for the work presented here.

The rest of the paper is organized as follows. Section \ref{sec:preliminaries} provides notation and
preliminary background material on the Bayesian formulation of Gaussian graphical model estimation. This section also briefly reviews the Bayesian graphical lasso estimator and its heuristic structure learning procedure presented in \cite{wang_2012_bayes}. In Section \ref{sec:bayesian_ridge}, the Bayesian graphical ridge-type estimators are proposed. Additionally, this section unpacks the latter's Bayesian block Gibbs sampler and computational insights associated with it. Finally, the Bayesian graphical ridge-type estimators are extended to the Bayesian adaptive graphical ridge-type estimators. Section \ref{sec:synthetic_examples} is dedicated to a synthetic study to compare the numerical accuracy between the Bayesian graphical lasso and Bayesian graphical ridge-type estimators. A cell signalling application is provided in Section \ref{sec:flow_cytometry} and concluding remarks in Section \ref{sec:discussion}.

\subsection{Contributions}{
In this paper a framework for a Bayesian graphical ridge-type inference is developed for low to moderate, $p \in \{10-100\}$, dimensionality, with $n > p$. The Bayesian graphical lasso, as well as its adaptive alternative by \cite{wang_2012_bayes} are reviewed and used as a departure point. Next, a Bayesian graphical ridge-type model is formulated followed by its corresponding block Gibbs sampler for sampling $\mathbf{\Omega}$. The block Gibbs sampler inherits all of the efficient properties from its influencer. The standard Bayesian ridge-type prior is extended to an adaptive Bayesian ridge-type estimator. Synthetic studies illustrate that the Bayesian graphical ridge-type model is a serious contender to its Bayesian graphical lasso counterpart. Lastly, an R package has been developed for the Bayesian graphical ridge-type block Gibbs sampler. The Markov Chain Monte Carlo (MCMC) sampler simulates precision matrices from the posterior distribution of the latter. The R package is available on The Comprehensive R Archive Network (CRAN)  \href{https://cran.r-project.org/web/packages/abglasso/index.html}{baygel}.\par
}

\section{Preliminaries}\label{sec:preliminaries}{

This section aims to introduce the required notation of undirected Gaussian graphical models; a comprehensive review is available in \cite{lauritzen1996graphical}. Let $\mathcal{G}=(\mathcal{V}, \mathcal{E})$ define an undirected graphical model where $\mathcal{V}=\{1,2,...,p\}$ is the set of nodes and $\mathcal{E} \subseteq \mathcal{V} \times \mathcal{V}$ the set of existing edges. The undirected graph provides a visual depiction of the conditional dependencies between the nodes. Following the notation used by \cite{mohammadi_2015_bayesian}, let 

\begin{equation*}
    \mathcal{M} = \{(i, j) \:|\: i, j \in \mathcal{V}, \:i < j\},
\end{equation*}

\noindent and $\Bar{\mathcal{E}} = \mathcal{W} \backslash \mathcal{E}$ denotes the set of non-existing edges. Next, define a zero mean Gaussian
graphical model with respect to the graph $\mathcal{G}$ as

\begin{equation*}
    \mathcal{W}_{\mathcal{G}} = \big\{\mathcal{N}_p\left(0,\mathbf{\Sigma}\right)\:|\: \mathbf{\Omega}=\mathbf{\Sigma}^{-1} \in \mathbb{M}^{+}\big\},
\end{equation*}

\noindent where $M^+$ is the space of positive definite matrices having entries $(i, j)=0$ whenever $(i, j) \in \Bar{E}$. Let the observations $\mathbf{x}=(\mathbf{x}_1, \mathbf{x}_2, ... ,\mathbf{x}_{n})$ be an independent and identically distributed sample from $\mathcal{W}_{\mathcal{G}}$. 

%
%
%

For the prior distribution of $\mathbf{\Omega}$, a brief review of \cite{wang_2012_bayes}'s Bayesian graphical lasso and Bayesian adaptive graphical lasso is provided. Recall that the object of the graphical lasso is to maximize the penalized log-likelihood

\begin{equation} \label{graphical_lasso}
   \argmax_{\mathbf{\Omega}\in \mathbb{M}^{+}} 
    \bigg\{\log(\mathrm{det}\mathbf{\Omega})-\mathrm{trace}(\frac{\mathbf{S}}{n}\mathbf{\Omega}) - \rho\Arrowvert \mathbf{\Omega} \Arrowvert_1\bigg\},
\end{equation}
  


\noindent here, $\rho \geq 0$ is the shrinkage parameter and $\mathbf{\Omega}=(\omega_{ij})$ is the precision matrix. The Bayesian graphical lasso prior is given by

\begin{equation} \label{bayes_glasso_prior}
    p\left(\mathbf{\Omega}\:|\lambda\right)=C^{-1}\prod_{i<j}\bigg\{\mathrm{DE}(\omega_{ij}\:|\:\lambda)\bigg\}
    \prod_{i=1}^{p}\bigg\{\mathrm{EXP}(\omega_{ii}\:|\:\lambda)\bigg\}\:\:\:\:(\mathbf{\Omega}\in \mathbb{M}^{+}).
\end{equation}

\noindent Here, the prior is given by the product of a double exponential (DE) with form $p(y)=\lambda/2\exp(-\lambda|y|)$ for the off diagonal elements and an exponential (EXP) with form $p(y)=\lambda \exp(-\lambda y)1_{y>0}$ for the diagonal. The mode of the posterior is  the graphical lasso estimate in \eqref{graphical_lasso} when $\rho=\lambda/n$, hence $\lambda$ can be viewed as a shrinkage lever. For computational simplicity a scale mixture of Gaussians is used to represent the double exponential distribution. However, this hierarchical restructuring requires additional simulation of the latent scale parameter in the proposed block Gibbs sampler.

To address the shortcomings of the double exponential, whereby it may over (under) shrink large (small) coefficients, the prior in \eqref{bayes_glasso_prior} is extended by allowing different shrinkage parameters, $\lambda_{ij}$, for each off-diagonal element $\omega_{ij}$. This formulation, namely the Bayesian adaptive graphical lasso (BAGLASSO) is given by

\begin{equation*}
    p(\mathbf{\Omega} \:|\: \{\lambda_{ij}\}_{i\leq j}) = 
    C^{-1}_{\{\lambda_{ij}\}_{i\leq j}}
    \prod_{i<j}\bigg\{\mathrm{DE}(\omega_{ij}\:|\:\lambda_{ij})\bigg\}
    \prod_{i=1}^{p}\bigg\{\mathrm{EXP}(\omega_{ii}\:|\:\frac{\lambda_{ii}}{2})\bigg\}\:\:\:\:(\mathbf{\Omega}\in \mathbb{M}^{+}),
\end{equation*}

\begin{equation} \label{unique_lambda_hier_glasso}
    p(\{\lambda_{ij}\}_{i<j}\:|\:\{\lambda_{ii}\}_{i=1}^p) \propto
    C_{\{\lambda_{ij}\}_{i\leq j}}
    \prod_{i<j}\mathrm{GA}(r,s).
\end{equation}

\noindent The BAGLASSO automatically controls the amount of shrinkage based on the value of $\omega_{ij}$. To see this, consider the distribution of each shrinkage parameter $\lambda_{ij}$ conditioned on $\mathbf{\Omega}$ in \eqref{unique_lambda_hier_glasso}

\begin{equation*}
    \lambda_{ij}\:|\:\mathbf{\Omega} \sim \mathrm{GA}(1+r,\:|\omega_{ij}|+s), 
\end{equation*}

\noindent implying that the conditional expectation of $\lambda_{ij}$ is $(1+r)/(|\omega_{ij}|+s)$. To this end, the amount of shrinkage applied to $\lambda_{ij}$ is inversely proportional to the value of $\omega_{ij}$, however, the hyperparameters $r$ and $s$ need to be sufficiently small to enjoy accurate adaptiveness.

The Bayesian graphical lasso, as well as its adaptive variant are not capable of producing $\omega_{ij}=0$ for $i \neq j$ since it places zero probability on these events. Graphical model determination under these priors can only be achieved via a heuristic procedure such as the thresholding approach recommended by \cite{carvalho2010horseshoe}. In particular, \cite{wang_2012_bayes} claims ${\omega_{ij}=0}$ if and only if

\begin{equation} \label{threshold_rule_wang}
   \frac{\Tilde{\rho}_{ij}}{E_g(\rho_{ij}\:|\:\mathbf{Y})} > 0.5.
\end{equation}

\noindent Here, $\Tilde{\rho}_{ij}$ is the mean estimate of the posterior partial correlation associated with the graphical lasso priors in \eqref{bayes_glasso_prior} and $g$ is chosen as standard conjugate Wishart $\mathrm{W}(3,\mathbf{I}_p)$ with parameter values selected based on evidence provided by \cite{jones2005experiments}.
}

\section{A Bayesian graphical ridge approach}\label{sec:bayesian_ridge}

\subsection{The graphical ridge-type prior}

Consider the penalized estimation problem in \eqref{graphical_lasso} as a departure point for precision matrix estimation. Using a ridge constraint in place of the $\ell_1$ results in a graphical ridge-type problem where the objective is to maximise the log-likelihood

\begin{equation}  \label{eq:graphical_ridge}
   \argmax_{\mathbf{\Omega}\in \mathbb{M}^{+}} 
    \bigg\{
    \log(\mathrm{det}\mathbf{\Omega})-\mathrm{trace}(\frac{\mathbf{S}}{n}\mathbf{\Omega}) - \frac{\rho}{2}\Arrowvert \mathbf{\Omega} \Arrowvert_2^2,
    \bigg\},
\end{equation}



\noindent over the space of positive definite matrices $M^+$. Here, $\rho>0$ is the shrinkage parameter and $\Arrowvert \mathbf{\Omega} \Arrowvert_2^2=\sum_{i<j}\sum_{i=1}^{p}\omega_{ij}^2$. Moreover, \eqref{eq:graphical_ridge} is a convex objective function and the Bayesian estimator is given by the maximum a posteriori (MAP) estimation

\begin{equation} \label{eq:bayes_gridge_prior1}
    p\left(\mathbf{\Omega}\:|\mu\:,\sigma\right)=C^{-1}\prod_{i<j}\bigg\{\mathrm{N}(\omega_{ij}\:|\:\mu=0\:,\sigma)\bigg\}
    \prod_{i=1}^{p}\bigg\{\mathrm{TN}(\omega_{ii}\:|\:\sigma)\bigg\}\:\:\:\:(\mathbf{\Omega}\in \mathbb{M}^{+}),
\end{equation}

\noindent where, $\mathrm{N}(x\:|\sigma)$ represents a Gaussian density function with form $p(y)=(\sigma\sqrt{2\pi})^{-1}\exp(-0.5[y/\sigma]^2)$ and $\mathrm{TN}(y\:|\sigma)$ represents a univariate left truncated at zero Gaussian density function with form  $p(y)=\sqrt{2}(\sigma\sqrt{\pi})^{-1}\exp(-0.5[y/\sigma]^2)1_{y>0}$. Moreover, $C$ is the normalising constant not involving $\sigma$


\begin{equation*}
    \begin{aligned}
        C &= \int_{\mathbf{\Omega}\in \mathbb{M}^{+}}
        \prod_{i<j}\bigg\{\mathrm{N}(\omega_{ij}\:|\:\mu=0\:,\sigma)\bigg\}
        \prod_{i=1}^{p}\bigg\{\mathrm{TN}(\omega_{ii}\:|\:\mu=0\:,\:\sigma)\bigg\}d\mathbf{\Omega}\\
        &=\int_{\mathbf{\tilde{\Omega}}\in \mathbb{M}^{+}}
        \prod_{i<j}\bigg\{\mathrm{N}(\tilde{\omega}_{ij}\:|\:\mu=0\:,1)\bigg\}
        \prod_{i=1}^{p}\bigg\{\mathrm{TN}(\Omega_{ii}\:|\:\mu=0\:,\:1)\bigg\}d\mathbf{\tilde{\Omega}}.
    \end{aligned}
\end{equation*}

\begin{figure}

\begin{minipage}{.5\linewidth}
\centering
\subfloat[]{\label{fig:marginal_density_omega_11}\includegraphics[scale=.28]{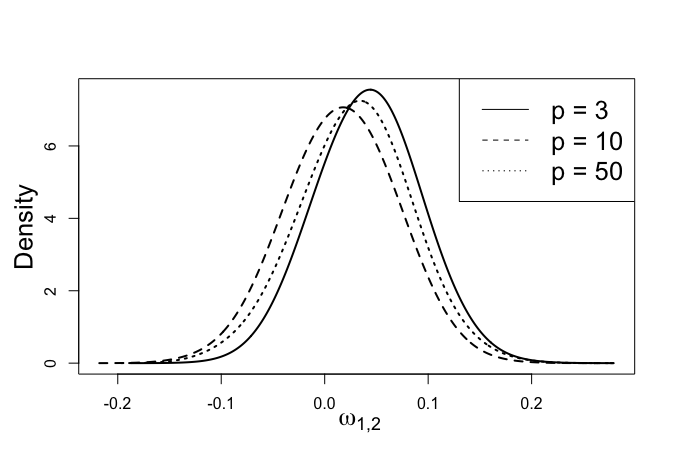}}
\end{minipage}%
\begin{minipage}{.5\linewidth}
\centering
\subfloat[]{\label{fig:tmarginal_density_omega_12}\includegraphics[scale=.28]{img/marginal_distributions/omega_12_png.png}}
\end{minipage}\\[-4ex]\par\medskip
\centering
\subfloat[]{\label{fig:marginal_density_rho_12}\includegraphics[scale=.28]{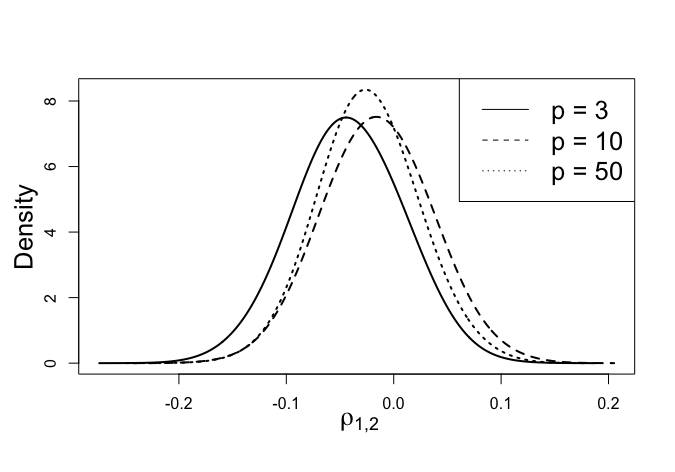}}

\caption{Marginal densities for diagonal (a), off-diagonal (b), and partial correlation (c) of $\mathbf{\Omega}$ when $\sigma=2$ and $p = 3$ (solid), 10 (dashed) and 50 (dotted).}
\label{fig:marginal_densities}
\end{figure}

\noindent The last equality relies on the substitution $\mathbf{\tilde{\Omega}} = \mathbf{\Omega}/\sigma$, and it holds that  $\{\mathbf{\tilde{\Omega}} :\mathbf{\tilde{\Omega}}\in \mathbb{M}^{+}\}=\{\mathbf{\Omega}:\mathbf{\Omega}\in \mathbb{M}^{+}\}$ for $\sigma>0$. The log of the posterior density is given by

\begin{equation*}
    \begin{aligned}
         l_r(\mathbf{\Omega}) &\propto 
         \log(\mathrm{det}\mathbf{\Omega})-\mathrm{trace}\left(\frac{\mathbf{S}}{n}\mathbf{\Omega}\right)
         +\frac{1}{n\sigma^2}\bigg\{\sum_{i=1}^p\sum_{j=1}^p\omega_{ij}^2\bigg\},
    \end{aligned}
\end{equation*}

\noindent where $l_r(\cdot)$ denotes a graphical ridge-type log-likelihood. The MAP estimate of $\mathbf{\Omega}$ which maximises this log-likelihood is given by

\begin{equation*}
    \underset{\mathbf{\Omega}\in \mathbb{M}^{+}}{\operatorname{argmax}} 
    \bigg\{       
    \log(\mathrm{det}\mathbf{\Omega})-\mathrm{trace}\left(\frac{\mathbf{S}}{n}\mathbf{\Omega}\right)
         +\frac{1}{n\sigma^2}\Arrowvert \mathbf{\Omega} \Arrowvert_2^2 
    \bigg\},
\end{equation*}

\noindent and the posterior mode of $\mathbf{\Omega}$ is given by \eqref{eq:graphical_ridge} with $\rho=2/n\sigma^2$. Figures (\ref{fig:marginal_density_omega_11}) - (\ref{fig:marginal_density_rho_12}) display the marginal distributions of one of the $p$ diagonal elements, one of the $p(p − 1)/2$ off-diagonal elements, and one of the $p(p − 1)/2$ partial correlations, respectively, when $\sigma=1$  and $p \in \{3, 10, 50\}$. The densities are based on the synthetic samples generated by the Monte Carlo sampling mechanism in Section \ref{subsec:block_gibbs_smapler}. Interestingly, the marginal distribution of the individual diagonal elements is not half Gaussian and tends to have an increasing mean and variance as $p$ increases. The partial correlation seems to settle down near zero with decreasing variation. This implies that the Bayesian graphical ridge-type prior also enjoys the desired property of partial correlations favoring near zero values as $p$ increases. It is for this reason that the Bayesian graphical lasso is used as a departure point, however, this desired characteristic may not be as pronounced in the Bayesian graphical ridge-type prior.  

\subsection{A data driven block Gibbs sampler}\label{subsec:block_gibbs_smapler}
The block Gibbs sampler described in \cite{wang_2012_bayes} serves as the impetus for the sampler described here. The target distribution associated with the prior in \eqref{eq:bayes_gridge_prior1} can be described as:

\begin{equation} \label{eq:target_dist}
    \begin{aligned}
        p(\mathbf{\Omega}\:|\:\mathbf{Y},\lambda)
        \propto
        \mathrm{det}\mathbf{\Omega}^{\frac{n}{2}}
        \exp\{-\mathrm{trace}(\frac{1}{2}\mathbf{S}\mathbf{\Omega})\}
        \prod_{i<j}\bigg\{\exp(-\frac{\omega_{ij}^2}{2\sigma^2})\bigg\}
        \prod_{i=1}^{p}\bigg\{\exp(-\frac{\omega_{ii}^2}{2\sigma^2})\bigg\}1_{\mathbf{\Omega}\in \mathbb{M}^{+}}.
    \end{aligned}
\end{equation}

\noindent It is clear that marginally, \eqref{eq:bayes_gridge_prior1} is maintained. It should be noted that positive definite constraint is only applicable to the elements of $\mathbf{\Omega}$. Furthermore, in contrast to the Bayesian graphical lasso target distribution, \eqref{eq:target_dist} does not include an unknown latent scale parameter. This light weight formulation may aid in reducing computational speed and complexity, albeit marginally. 

The block Gibbs sampler described here illustrates how to update $\mathbf{\Omega}$ one column and row at a time. Without loss of generality, consider the last column and row and let $\mathbf{\Psi}$ be a symmetric $p \times p$ matrix with a zero main diagonal and $\boldsymbol{\tau}$ in the upper and lower off-diagonal entries. Here, $\boldsymbol{\tau}$ represents a vector of $\sigma^{2}$ values. Partition the matrices $\mathbf{\Theta}$, $\mathbf{S}$ and $\mathbf{\Psi}$ as follows

\begin{equation} \label{eq:partition_gibbs}
    \mathbf{\Omega}=
        \begin{pmatrix}
         \mathbf{\Omega}_{11} & \boldsymbol{\omega}_{12}\\
         \boldsymbol{\omega}_{21} & \omega_{22}
        \end{pmatrix},
    \:\:\:\:
    \mathbf{S} = 
        \begin{pmatrix}
         \mathbf{S}_{11} & \mathbf{s}_{12}\\
         \mathbf{s}_{21} & s_{22}
        \end{pmatrix},
    \:\:\:\:
    \mathbf{\Psi} = 
        \begin{pmatrix}
         \mathbf{\Psi_{11}} & \boldsymbol{\sigma}\\
         \boldsymbol{\sigma} & 0
        \end{pmatrix}.
\end{equation}

\noindent Recall that the $\mathrm{det}\mathbf{\Omega}^{\frac{n}{2}}$ can be represented as

\begin{equation*}
    \begin{aligned}
        \mathrm{det}\mathbf{\Omega}^{\frac{n}{2}} = (\omega_{22}-\boldsymbol{\omega}_{21}\boldsymbol{\Omega}^{-1}_{11}\boldsymbol{\omega}_{12})^{\frac{n}{2}}\mathrm{det}\mathbf{\Omega_{11}}^{\frac{n}{2}}
        &\propto
        (\omega_{22}-\boldsymbol{\omega}_{21}\boldsymbol{\Omega}^{-1}_{11}\boldsymbol{\omega}_{12})^{\frac{n}{2}},
    \end{aligned}
\end{equation*}

\noindent since we are only interested in the last column and row. Similarly,
\begin{equation*}
    \mathrm{trace}(\frac{1}{2}\mathbf{S}\mathbf{\Omega})\propto -\frac{1}{2}(2\boldsymbol{s_{21}}\boldsymbol{\omega}_{12} + s_{22}\omega_{22}).
\end{equation*}

\noindent Moreover, 

\begin{equation*}
    \begin{aligned}
        &\prod_{i<j}\bigg\{\mathrm{exp}\left(-\frac{1}{2}\left(\frac{\omega_{ij}}{\sigma}\right)^2\right)\bigg\} 
        \mathrm{exp}\left(-\frac{1}{2}\left(\frac{\omega_{22}}{\sigma}\right)^2\right)\\
       &=\mathrm{exp}\bigg\{-\frac{1}{2}\left(\boldsymbol{\omega_{12}}^{\top}\boldsymbol{D}_{\boldsymbol{\tau}}^{-1}\boldsymbol{\omega_{12}}+\left(\frac{\omega_{22}}{\sigma}\right)^2\right)\bigg\}; \:\:\:\: \mathrm{for} \:\: \omega_{22} > 0,
    \end{aligned}
\end{equation*}

       

\noindent where $\boldsymbol{D}_{\boldsymbol{\tau}}= \mathrm{diag}(\boldsymbol{\tau})$. The conditional distribution of the last column in $\boldsymbol{\Theta}$ is
\begin{equation*}
    \begin{aligned}
        p(\boldsymbol{\omega_{12}},\omega_{22}\:&|\: \boldsymbol{\Theta}_{11},\boldsymbol{\Psi},\boldsymbol{S},\sigma)
        \propto
        \left(\omega_{22}-\boldsymbol{\omega}_{21}\boldsymbol{\Omega}^{-1}_{11}\boldsymbol{\omega}_{12}\right)^{\frac{n}{2}}\\
        &\times 
        \mathrm{exp}\left[-\frac{1}{2}\bigg\{\boldsymbol{\omega}_{21}\boldsymbol{D}_{\boldsymbol{\tau}}^{-1}\boldsymbol{\omega}_{12}  +2\boldsymbol{s}_{21}\boldsymbol{\omega}_{12} + \omega_{22}\left(s_{22}+\frac{\omega_{22}}{\sigma^2}\right)\bigg\}\right]\\
        &= 
        \left(\omega_{22}-\boldsymbol{\theta}_{21}\boldsymbol{\Theta}^{-1}_{11}\boldsymbol{\theta}_{12}\right)^{\frac{n}{2}}\\
        &\times 
        \mathrm{exp}\left[-\frac{1}{2}\bigg\{\boldsymbol{\omega}_{21}\boldsymbol{D}_{\boldsymbol{\tau}}^{-1}\boldsymbol{\omega}_{12}  +2\boldsymbol{s}_{21}\boldsymbol{\omega}_{12} + \omega_{22}\left(\sum_{k=0}^{1}{1 \choose k} s_{22}^{1-k}+\left(\frac{\omega_{22}}{\sigma^2}\right)^k\right)\bigg\}\right]\\
       &\propto
       \left(\omega_{22}-\boldsymbol{\theta}_{21}\boldsymbol{\Theta}^{-1}_{11}\boldsymbol{\theta}_{12}\right)^{\frac{n}{2}}\\
       &\times 
       \mathrm{exp}\left[-\frac{1}{2}\bigg\{\boldsymbol{\omega}_{21}\boldsymbol{D}_{\boldsymbol{\tau}}^{-1}\boldsymbol{\theta}_{12}  +2\boldsymbol{s}_{21}\boldsymbol{\omega}_{12} + \theta_{22}\left(s_{22}+1\right)\bigg\}\right]; \:\: \mathrm{for} \:\: k=0.\\
    \end{aligned}
\end{equation*}

\noindent Consider the following change of variables

\begin{equation*}
    \begin{aligned}
        \boldsymbol{\beta} &=\boldsymbol{\omega}_{12}\\
        \gamma &= \omega_{22}-\boldsymbol{\omega}_{21}\boldsymbol{\Omega}^{-1}_{11}\boldsymbol{\omega}_{12},
    \end{aligned}
\end{equation*}

\noindent with the Jacobian independent of $(\boldsymbol{\beta},\gamma)$, yields the following conditional distribution

\begin{equation*}
    \begin{aligned}
        p(\boldsymbol{\beta},\gamma\:|\: \boldsymbol{\Omega}_{11},\boldsymbol{\Psi},\boldsymbol{S},\sigma)
    \propto
    &\gamma^{\frac{n}{2}}\mathrm{exp}\left(-\frac{s_{22}+1}{2}\gamma\right)\\
    &\times
    \mathrm{exp}\left[-\frac{1}{2}\left(\boldsymbol{\beta}^{\top}\{\boldsymbol{D}_{\boldsymbol{\tau}}^{-1}+ (s_{22} +1)\boldsymbol{\Theta}_{11}^{-1} \}\boldsymbol{\beta}+2\boldsymbol{s}_{21}\boldsymbol{\beta}\right)\right].\\
    \end{aligned}
\end{equation*}
\noindent It follows that
\begin{equation*}
    (\gamma,\boldsymbol{\beta}\:|\: \boldsymbol{\Theta}_{11},\boldsymbol{\Psi},\boldsymbol{S},\sigma) 
    \sim \mathrm{GA}\left(\frac{n}{2}+1,\frac{s_{22}+1}{2}\right)
    \mathrm{N}\left(-\boldsymbol{C}\boldsymbol{s}_{21},\boldsymbol{C}\right),
\end{equation*}

\noindent where $\mathrm{GA}(\alpha,\eta)$ represents a gamma distribution with shape parameter $\alpha$ and scale parameter $\eta$ and $\boldsymbol{C}=\bigg\{(s_{22}+1)\boldsymbol{\Omega}_{11}+\boldsymbol{D}_{\boldsymbol{\tau}}^{-1}\bigg\}^{-1}$. The block Gibbs sampler maintains the positive definite constraint on $\mathbf{\Omega}$ and readers are referred to the rigorous explanation provided for the Bayesian graphical lasso sampler for additional insights. Finally, The block Gibbs sampler can be summarised as follows

\begin{algorithm}
	\caption{Block Gibbs sampler}\label{alg:block_gibbs_sampler} 
	
	\begin{algorithmic}[]
	\Require An initial estimate of $\mathbf{\Omega} \in \mathbb{M}^{+}.$
		\For {$i=1,\ldots,p$}
		    \State 1) Partition $\mathbf{\Omega}$, $\mathbf{S}$ and $\mathbf{\Psi}$ as in \eqref{eq:partition_gibbs}.
		    \State 2) Sample $\gamma \sim \mathrm{GA}(n/2+1,\:(s_{22}+1)/2)$ and $\mathbf{\beta} \sim \mathrm{N}\left(-\boldsymbol{C}\boldsymbol{s}_{21},\boldsymbol{C}\right)$.
		    \State 3) Update $\mathbf{\omega}_{12} = \mathbf{\beta}$, $\mathbf{\omega}_{21} = \mathbf{\beta}^{\top}$ and $\omega_{22} = \gamma + \mathbf{\beta}^{\top}\mathbf{\Omega_{11}^{-1}}\mathbf{\beta}$.
		\EndFor
	\end{algorithmic} 
\end{algorithm}
  
\subsection{Computational speed investigation}\label{subsec:computational_speed_study}

This section aims to compare and evaluate the computational speed and scalability of the block Gibbs sampler in Algorithm \ref{alg:block_gibbs_sampler}. The synthetic samples generated by the Monte Carlo sampling mechanism in Section \ref{subsec:block_gibbs_smapler} were used where the number of observations $n$ were set to 500 and $\mathbf{\Omega}$ was initialised at the identity matrix. The computations were evaluated on a 2 GHz Quad-Core Intel Core i5 running macOS Monterey (version 12.4) using the R (version 4.2.0) programming language.

\begin{figure}[H]

\begin{minipage}{.5\linewidth}
\centering
\subfloat[]{\label{fig:computation_time}\includegraphics[scale=.28]{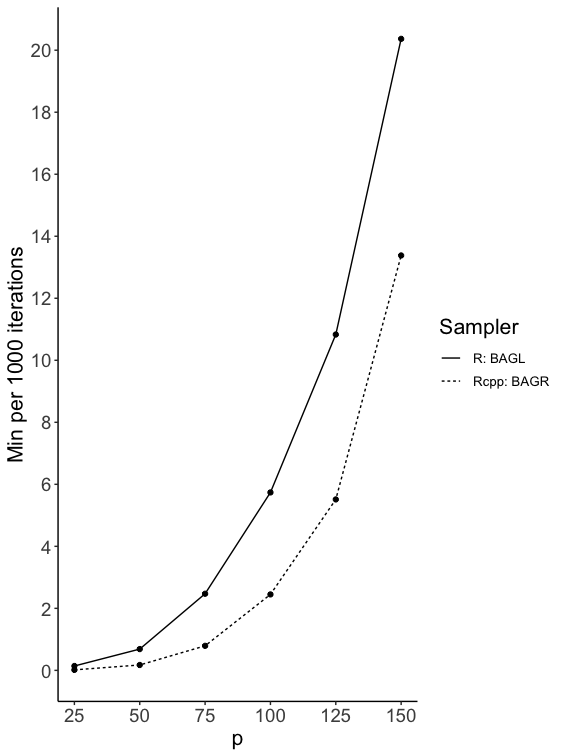}}
\end{minipage}%
\begin{minipage}{.5\linewidth}
\centering
\subfloat[]{\label{fig:proportional_computation_impv}\includegraphics[scale=.28]{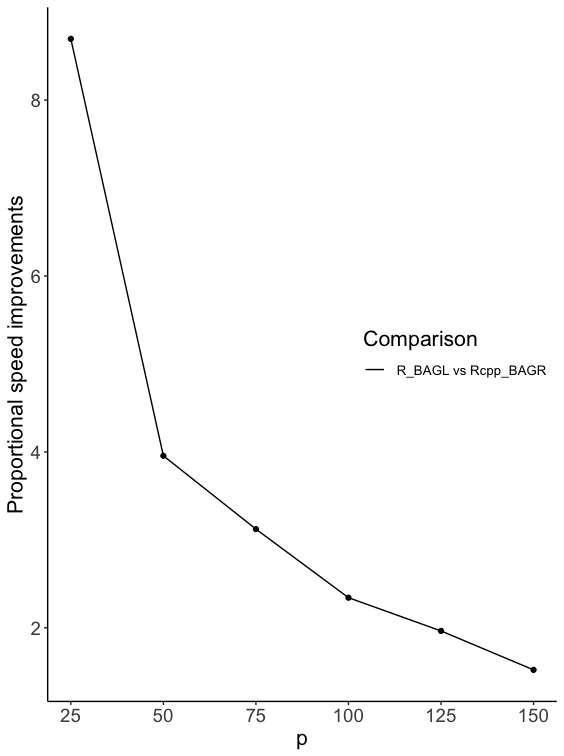}}
\end{minipage}\\[-4ex]\par\medskip

\caption{Computational time comparisons as a function of $p$.}

\label{fig:computational_speed}
\end{figure}

Computational insights for the block Gibbs samplers associated with the Bayesian adaptive graphical lasso and the Bayesian standard graphical ridge-type ($\sigma = 1$), referred to as "BAGL" and "BSGR" respectively, are displayed in Figure \ref{fig:computational_speed}. Figure \ref{fig:computation_time} compares the number of minutes required to compute 1000 iterations, as $p$ varies, between BSGR and BAGL. Here, a single iteration constitutes the block Gibbs sampler to update each entry of $\mathbf{\Omega}$. The BSGR is implemented with C++ code linked to R via the 'Baygel' R package. The BAGL is implemented using base R from the 'abglasso' R package. The C++ BSGR algorithm took approximately 2.4 minutes to generate 1000 iterations for $p=100$, whereas the R BAGL algorthim took 5.8 minutes. Figure \ref{fig:proportional_computation_impv} highlights the proportional speed improvements between the native R and C++ based samplers. The C++ based algorithms are noticeably faster than the native R implementations, however, the proportional speed improvements display exponential decay as $p$ increases. This is not surprising given that the samplers compute inverse of matrices, as well as Cholesky decompositions in order to sample from multivariate Gaussian distributions. In other words, the block Gibbs samplers have at least $O(n^3)$ complexity. In this regard, matrix factorisations are commonly carried out using external software libraries such as the Linear Algebra Package (LAPACK) or the Basic Linear Algebra Subprograms (BLAS). Both C++ and R, typically make use of LAPACK or BLAS or similar for numerical linear algebra calculations implying that the former's block Gibbs samplers will experience diminishing speed gains compared to the latter as $p$ increases. Finally, the convergence of the BSGR block Gibbs sampler was assessed using the inefficiency factor  \cite{kim1998stochastic} $1+2\sum_{i=1}^{\infty}\eta(k)$, where $\eta(k)$ is the sample autocorrelation at lag $k$. The process entailed using 3000 samples after 1000 burn-in iterations and 300 lags, of which the median inefficiency factor was calculated. This procedure was repeated 100 times and the median of the median inefficiency factors among all of the elements of $\mathbf{\Omega}$ was 0.76. This suggests that Markov chain Monte Carlo (MCMC) process mixes well.

\subsection{Bayesian adaptive graphical ridge}\label{sec:bayesian_adaptive_graphical_ridge}
The Bayesian graphical ridge-type prior in \eqref{eq:bayes_gridge_prior1} requires the selection of $\sigma$. The structure of the prior and the block Gibbs sampler in Section \ref{subsec:block_gibbs_smapler} allows for extended hierarchical formulations of the prior to allow for adaptive shrinkage on different off-diagonal elements of $\mathbf{\mathbf{\Omega}}$. That is, the Bayesian framework facilitates the usage of a hyperprior on $\tau = \sigma^{2}$ and extending the block Gibbs sampler to sample from the posterior distribution thereof. Consider the following Bayesian adaptive graphical ridge-type set of priors

\begin{equation}\label{eq:bayesian_adaptive_ridge_prior}
    \begin{aligned}
        p\left(\mathbf{y_i}\:|\mathbf{\Omega} \right)&=N_p\left(\mathbf{y}_i\:|0,\:\mathbf{\Omega}^{-1}\right) \:\:\: (i=1,\cdots,n)\\
        p\left(\mathbf{\Omega}\:|\mu\:,\sigma\right)&=C^{-1}_{\{\sigma_{ij}\}_{i<j}}\prod_{i<j}\bigg\{\mathrm{N}(\omega_{ij}\:|\:\mu=0\:,\sigma_{ij})\bigg\}
    \prod_{i=1}^{p}\bigg\{\mathrm{TN}(\omega_{ii}\:|\:\mu=0,\sigma_{ii})\bigg\}1_{\mathbf{\Omega}\in \mathbb{M}^{+}}\\
        p\left(\{\tau_{ij}\}_{i<j}\:|\{\mu_{ii},\sigma_{ii}\}^{p}_{i=1} \right)& \propto C_{\{\tau_{ij}\}_{i<j}}\prod_{i<j}\mathrm{IGA}(a,b),
    \end{aligned}
\end{equation}

\noindent where $\mathrm{IGA}(a,b)$ represents an inverse gamma distribution with a shape parameter $a$ and scale parameter $b$ and form $p(y)=b^a/\Gamma(a)y^{a+1}\mathrm{exp}(-b/y)$. Moreover, $C_{\{\tau_{ij}\}_{i<j}}$ is the intractable normalisation constant and $\sigma_{ii}$ and $\mu_{ii}$ are hyperparameters for the main diagonal elements. Notice that the terms $C_{\{\sigma_{ij}\}_{i<j}}$ in \eqref{eq:bayesian_adaptive_ridge_prior} cancel out when sampling from the posterior of $\tau_{ij}$. That being said, model fitting, using the block Gibbs sampler in Section \ref{subsec:block_gibbs_smapler} is straight forward and a simple addition of updating different $\tau_{ij}$ from different inverse gamma distributions will suffice. Conditional on $\boldsymbol{\Omega}$, the posterior is given by  

\begin{equation}\label{eq:posterior_sig_ij}
    \tau_{ij}\:|\:\mathbf{\Omega} \sim \mathrm{IGA}(a+\frac{1}{2},b+\frac{\omega_{ij}^2}{2}).
\end{equation}

\section{Synthetic examples}\label{sec:synthetic_examples}
The synthetic study design aims to assess the parameter estimation performance of the BAGL \eqref{unique_lambda_hier_glasso}, the BSGR \eqref{eq:bayes_gridge_prior1} and the Bayesian adaptive graphical ridge "BAGR" \eqref{eq:bayesian_adaptive_ridge_prior}. The parameters of the gamma priors in \eqref{unique_lambda_hier_glasso} are set to $r = 10^{−2}$ and $s = 10^{−6}$, as proposed. Additionally, $\lambda_{ii}=1$ for $i=1, \cdots, p$. The parameters of the inverse gamma priors in \eqref{eq:bayesian_adaptive_ridge_prior} are set to  $a = 1$ and $b = 10^{−2}$. These parameters were selected using a grid search approach over all the models defined in Tab and  $\sigma_{ii}=1$ for $i=1, \cdots, p$. The basic assumptions for all simulations is that $\mathbf{x}_1, \mathbf{x}_2, ... ,\mathbf{x}_{n}$ are generated from a Gaussian $N_p(0,\mathbf{\Sigma})$ where the true precision is given by $\mathbf{\Omega}=\mathbf{\Sigma}^{-1}$. Structure learning is omitted from the investigation given that the focus of the graphical ridge approach is on accuracy and not sparsity. That being said, if structure learning is required, both Bayesian techniques described here require a heuristic treatment and techniques described in Section \ref{sec:preliminaries} may be used. The following 6 models are considered.

\begin{table}
\centering
\begin{tabular}{p{0.08\linewidth} | p{0.12\linewidth} | p{0.7 \linewidth}}
\hline
\textbf{Model} & \textbf{Type}     & \textbf{Component}  \\ 
\hline
M1     & Diagonal & $\omega_{ii} \sim \mathrm{U(1,\:1.25)}$, where $\mathrm{U(\cdot)}$ denotes a uniform distribution.       \\
      &          &            \\
M2     & AR(1)    & $\omega_{ij}=0.99^{|i-j|}$.       \\
      &          &            \\
M3     & AR($p-3$)  & $\omega_{ii}=3$ and $\omega_{p-3,p}=\omega_{p,p-3}=0.1$ and the sequential decay from $\omega_{ii}$ to $\omega_{p-3,i}$ and $\omega_{i,p-3}$ is $2.9/p$ in magnitude.       \\
      &          &            \\
M4     & Cluster  & $\omega_{ii}=2,\omega_{ij}=1$ for $1 \leq i \neq j \leq p/2$, $\omega_{ij}=1$ for $p/2+1 \leq i \neq j \leq p$ and $\omega_{ij}=0$ otherwise.       \\
      &          &            \\
M5     & Cluster  & $\omega_{ii}=2,\omega_{ij}=1$ for $1 \leq i \neq j \leq p/5$, $\omega_{ij}=1$ for $p/5+1 \leq i \neq j \leq p$ and $\omega_{ij}=0$ otherwise.       \\
      &          &            \\
M6     & Full     & $\omega_{ii}=2$ and $\omega_{ij}=1$.       \\
\hline
\end{tabular}
\caption{Precision matrix structures and element compositions used in the synthetic examples.}
\label{tab:syn_studies_models}
\end{table}


\noindent The sample sizes and dimensions for each model are $n=500$ and $p \in\{10,30,50\}$, respectively. The estimates are based on $5000$ Monte Carlo iterations after $2000$ burn-in iterations. 
The approach by \cite{smith2022empowering} is followed to assess the performance of the precision matrix estimation. In particular, seven loss functions are considered and defined in Table \ref{tab:syn_loss_functions}, where $p$ denotes the dimension and $\gamma_i$ the $i^{th}$ eigenvalue, respectively. Tables \ref{tab:syn_loss_results_med} and \ref{tab:syn_loss_results_sd} report the median and standard error, respectively, of L1, L2, EL1, EL2, MAXEL1, MINEL1 and STEIN for $p = 10,\:30,\:50$ in models $1-6$ based on $30$ replications. For each scenario, the best performing measure is boldfaced.

\begin{table}[h]
\centering
\begin{tabular}{lll}
\hline                              

Measure                               & Loss function & Abbreviation  \\
\hline
Matrix $L_1$-norm                     & $\Arrowvert \hat{\mathbf{\Omega}} -  \mathbf{\Omega}\Arrowvert_1 = max_{1 \leq j \leq p}\sum_{i=1}^p|\hat{\Omega}_{ij}-\Omega_{ij}|$              & L1              \\

Frobenius loss                        &  $\Arrowvert \hat{\mathbf{\Omega}} -  \mathbf{\Omega}\Arrowvert_F$, where $\Arrowvert A\Arrowvert_F^2=\mathrm{trace}(AA^{\top})$             &  L2             \\

$L_1$ eigenvalue loss                 & $\sum_{i=1}^p|\hat{\gamma}_i-\gamma_i|/p$              &   EL1             \\
$L_2$ eigenvalue loss                  & $\sum_{i=1}^p(\hat{\gamma}_i-\gamma_i)^2/p$              & EL2              \\
$L_1$ loss on the largest eigenvalue  & $|\hat{\gamma}_{max}-\gamma_{max}|$              & MAXEL1              \\
$L_1$ loss on the smallest eigenvalue & $|\hat{\gamma}_{min}-\gamma_{min}|$              &  MINEL1          \\   
Stein's loss                         & $\mathrm{trace}(\hat{\Sigma}\Sigma^{-1})-\mathrm{log}(|\hat{\Sigma}\Sigma^{-1}|)-p$              &  STEIN         \\  
\hline
\end{tabular}
\caption{The loss functions used in the synthetic studies to assess the numerical accuracy of the BAGL, BSGR and BAGR estimates.}
\label{tab:syn_loss_functions}
\end{table}

\noindent The results from Tables \ref{tab:syn_loss_results_med} and \ref{tab:syn_loss_results_sd} provide interesting insight into the behaviour of the Bayesian graphical model estimation techniques. First, the BAGL estimator consistently outperforms the BAGR and BSGR estimators across all loss functions for model M1 for $p \in {10,30,50}$ and for M4 for $p=10$.  The former observation is not surprising given the sparse nature of the diagonal model. Second, the BAGR estimator performs remarkably well in the the remaining models across all loss functions, especially in model M3; the BSGR estimator being the runner up. Third, the standard errors of the BAGL and BSGR estimators remain relatively constant throughout the dimension spectrum. The BAGR displays larger standard errors for increasing $p$ across the non-sparse structures, however, its loss values are significantly lower compared to the others.

\begin{sidewaystable}[!ht]
\centering
\noindent\setlength\tabcolsep{4pt}%
\scriptsize
\begin{tabular}{lllllllllllllllllllll} 
\hline
\multicolumn{1}{c}{} & \multicolumn{3}{c}{\bf M1}  & \multicolumn{3}{c}{\bf M2} & \multicolumn{3}{c}{\bf M3} & \multicolumn{3}{c}{\bf M4} & \multicolumn{3}{c}{\bf M5}  & \multicolumn{3}{c}{\bf M6} \\ 
\cmidrule(lr){2-4}\cmidrule(lr){5-7}\cmidrule(lr){8-10}\cmidrule(lr){11-13}\cmidrule(lr){14-16}\cmidrule(lr){17-19}
\multicolumn{1}{c}{} & \multicolumn{1}{c}{BAGL} & \multicolumn{1}{c}{BAGR} & \multicolumn{1}{c}{BSGR} & \multicolumn{1}{c}{BAGL} & \multicolumn{1}{c}{BAGR} & \multicolumn{1}{c}{BSGR} & \multicolumn{1}{c}{BAGL} & \multicolumn{1}{c}{BAGR} & \multicolumn{1}{c}{BSGR} & \multicolumn{1}{c}{BAGL} & \multicolumn{1}{c}{BAGR} & \multicolumn{1}{c}{BSGR} & \multicolumn{1}{c}{BAGL} & \multicolumn{1}{c}{BAGR} & \multicolumn{1}{c}{BSGR} & \multicolumn{1}{c}{BAGL} & \multicolumn{1}{c}{BAGR} & \multicolumn{1}{c}{BSGR}  \\ 
\hline
\multicolumn{21}{c}{p=10} 
\\
\\
L1      & \textbf{0.13} & 0.51  & 0.50 
        & 1.63 & \textbf{0.50} & 1.19  
        & 2.63 & \textbf{1.74} & 4.17 
        & \textbf{0.79}  & 1.14 & 1.07   
        & 0.80 & 0.61 & \textbf{0.59} 
        & 2.83  & \textbf{1.29} & 2.20 \\

L2      & \textbf{0.19} & 0.45  & 0.45 
        & 1.54 & \textbf{0.41} & 1.11 
        & 2.08 & \textbf{1.45} & 3.57 
        & \textbf{0.78} & 0.97 & 0.93 
        & 0.63 & 0.49 & \textbf{0.48} 
        & 2.30  & \textbf{1.04} & 1.70 \\ 
        
EL1     & \textbf{0.03} & 0.09 & 0.09 
        & 0.16 & \textbf{0.04} & 0.11
        & 0.27 & \textbf{0.15} & 0.42 
        & \textbf{0.14} & 0.17 & 0.17 
        & 0.09 & \textbf{0.07} & \textbf{0.07} 
        & 0.33 & \textbf{0.19} & 0.26 \\
        
EL2     & 0.01 & 0.01 & 0.01
        & 0.24 & \textbf{0.02} & 0.12    
        & 0.36 & \textbf{0.10} & 0.16 
        & \textbf{0.04} & 0.06 & 0.05
        & 0.03 & \textbf{0.01} & \textbf{0.01} 
        & 0.49 & \textbf{0.06} & 0.24 \\
        
MAXEL1  & \textbf{0.07} & 0.23 & 0.23 
        & 1.54 & \textbf{0.40} & 1.11  
        & 1.83 & \textbf{0.58}  & 3.30 
        & \textbf{0.14} & 0.48 & 0.16 
        & 0.53 & 0.16 & \textbf{0.15} 
        & 2.14 & \textbf{0.66} & 1.46 \\
        
MINEL1  & \textbf{0.04} & 0.10 & 0.10 
        & 0.01 & 0.01 & 0.01 
        & 0.01 & 0.01  & 0.01 
        & \textbf{0.12} & 0.17 & 0.16 
        & \textbf{0.08} & 0.08 & 0.08 
        & 0.17 & 0.17 & \textbf{0.16} \\ 
        
STEIN  & \textbf{0.02} & 0.11 & 0.11 
        & 0.14 & \textbf{0.13} & \textbf{0.13} 
        & 0.13 & \textbf{0.11} & 0.14 
        & \textbf{0.06} & 0.11 & 0.11 
        & \textbf{0.09} & 0.11 & 0.11 
        & 0.14 & \textbf{0.11} & 0.12 \\ 
\\
\multicolumn{21}{c}{p=30}  
\\
\\
L1      & \textbf{0.24} & 1.58  & 1.57 
        & 27.60 & \textbf{2.41} & 11.86  
        & 62.18 & \textbf{6.12} & 43.76 
        & 8.50  & \textbf{5.29} & 5.32   
        & 11.98 & \textbf{2.58}  & 2.77 
        & 30.00  & \textbf{4.59} & 15.84 \\

L2      & \textbf{0.39} & 1.52  & 1.52 
        & 26.69 & \textbf{1.81} & 11.37 
        & 55.33 & \textbf{4.56} & 38.70 
        & 10.50 & \textbf{4.00} & 4.99 
        & 11.88 & \textbf{1.86} & 2.10 
        & 29.81  & \textbf{3.20} & 14.36 \\ 
        
EL1     & \textbf{0.03} & 0.18 & 0.18 
        & 0.92 & \textbf{0.07} & 0.39
        & 2.27 & \textbf{0.18} & 1.51 
        & 0.64 & \textbf{0.34} & 0.43 
        & 0.46 & \textbf{0.16} & 0.18 
        & 1.10 & \textbf{0.29} & 0.71 \\
        
EL2     & \textbf{0.01} & 0.06 & 0.06
        & 23.30 & \textbf{0.10} & 4.30    
        & 93.88 & \textbf{0.36} & 49.81 
        & 3.64 & \textbf{0.41} & 0.73
        & 3.07 & \textbf{0.07} & 0.11 
        & 29.32 & \textbf{0.18} & 6.80 \\
        
MAXEL1  & \textbf{0.11} & 0.60 & 0.60 
        & 26.43 & \textbf{1.70} & 11.35  
        & 51.00 & \textbf{2.21}  & 38.19 
        & 6.95 & 2.96 & \textbf{2.62} 
        & 9.17 & \textbf{1.15} & 1.64 
        & 29.65 & \textbf{1.70} & 14.19 \\
        
MINEL1  & \textbf{0.05} & 0.18 & 0.18 
        & 0.01 & 0.01 & 0.01 
        & 0.01 & 0.01  & 0.01 
        & \textbf{0.22} & 0.28 & 0.28 
        & \textbf{0.08} & 0.14 & 0.14 
        & \textbf{0.16} & 0.29 & 0.29 \\
        
STEIN  & \textbf{0.09} & 0.97 & 0.97 
        & 64.17 & \textbf{1.17} & 1.25 
        & 15.14 & \textbf{1.00}  & 1.92 
        & 0.99 & \textbf{0.98} & 1.02 
        & 16.25 & 0.97 & \textbf{0.96} 
        & 22.05 & \textbf{0.97} & 1.20 \\ 
\\
\multicolumn{21}{c}{p=50}                                                      \\&&&&&&&&&&&&&&&&&&&&\\
L1      & \textbf{0.33} & 2.86  & 2.84 
        & 43.96 & \textbf{4.92} & 25.31  
        & 109.03 & \textbf{12.24} & 87.81 
        & 25.06  & \textbf{11.29} & 12.57   
        & 20.02 & \textbf{6.31}  & 7.49 
        & 50.06  & \textbf{10.69} & 33.35 \\

L2      & \textbf{0.63} & 2.93  & 2.92 
        & 42.37 & \textbf{3.82} & 23.23 
        & 101.10 & \textbf{8.47} & 78.98 
        & 35.02 & \textbf{8.77} & 13.83 
        & 19.95 & \textbf{4.36} & 6.13 
        & 49.85  & \textbf{6.63} & 31.61 \\ 
        
EL1     & \textbf{0.04} & 0.27 & 0.27 
        & 0.94 & \textbf{0.10} & 0.51
        & 2.63 & \textbf{0.24} & 1.91 
        & 1.11 & \textbf{0.51} & 0.71
        & 0.47 & \textbf{0.25} & 0.29 
        & 1.09 & \textbf{0.43} & 0.97 \\
        
EL2     & \textbf{0.01} & 0.14 & 0.14
        & 35.24 & \textbf{0.27} & 11.73    
        & 195.36 & \textbf{0.79} & 124.67 
        & 24.18 & \textbf{1.27} & 3.69
        & 5.62 & \textbf{0.31} & 0.70 
        & 49.35 & \textbf{0.64} & 19.90 \\
        
MAXEL1  & \textbf{0.16} & 1.03 & 1.02 
        & 41.82 & \textbf{3.61} & 24.22  
        & 94.70 & \textbf{3.93}  & 77.59 
        & 24.55 & \textbf{6.53} & 8.94 
        & 16.04 & \textbf{3.38} & 5.60 
        & 49.66 & \textbf{4.73} & 31.39 \\
        
MINEL1  & \textbf{0.08} & 0.23 & 0.23 
        & 0.01 & 0.01 & 0.01 
        & 0.01 & 0.01  & 0.01 
        & \textbf{0.18} & 0.35 & 0.34 
        & \textbf{0.09} & 0.17 & 0.17 
        & \textbf{0.15} & 0.34 & 0.34 \\ 
        
STEIN  & \textbf{0.22} & 2.82 & 2.80 
        & 135.65 & \textbf{3.34} & 3.47 
        & 62.32 & \textbf{4.71}  & 5.99 
        & 34.66 & \textbf{2.85} & 3.03 
        & 31.10 & \textbf{2.83} & 2.87 
        & 40.01 & \textbf{2.81} & 3.44 \\

\hline
\end{tabular}
\caption{Summary of L1, L2, EL1, EL2, MAXEL1, MINEL1 and STEIN for a diagonal, AR($1$), AR($p-3$), a cluster model with two equally sized clusters, a cluster model with two clusters with a 1:5 size ratio and a full  model. The median loss values reported here are based on 30 replications for both the BAGL, BAGR and BSGR estimators. The best performing values are boldfaced.}
\label{tab:syn_loss_results_med}
\end{sidewaystable}

\clearpage

\begin{sidewaystable}[!ht]
\centering
\noindent\setlength\tabcolsep{4pt}%
\scriptsize
\begin{tabular}{lllllllllllllllllllll} 
\hline
\multicolumn{1}{c}{} & \multicolumn{3}{c}{\bf M1}  & \multicolumn{3}{c}{\bf M2} & \multicolumn{3}{c}{\bf M3} & \multicolumn{3}{c}{\bf M4} & \multicolumn{3}{c}{\bf M5}  & \multicolumn{3}{c}{\bf M6} \\ 
\cmidrule(lr){2-4}\cmidrule(lr){5-7}\cmidrule(lr){8-10}\cmidrule(lr){11-13}\cmidrule(lr){14-16}\cmidrule(lr){17-19}
\multicolumn{1}{c}{} & \multicolumn{1}{c}{BAGL} & \multicolumn{1}{c}{BAGR} & \multicolumn{1}{c}{BSGR} & \multicolumn{1}{c}{BAGL} & \multicolumn{1}{c}{BAGR} & \multicolumn{1}{c}{BSGR} & \multicolumn{1}{c}{BAGL} & \multicolumn{1}{c}{BAGR} & \multicolumn{1}{c}{BSGR} & \multicolumn{1}{c}{BAGL} & \multicolumn{1}{c}{BAGR} & \multicolumn{1}{c}{BSGR} & \multicolumn{1}{c}{BAGL} & \multicolumn{1}{c}{BAGR} & \multicolumn{1}{c}{BSGR} & \multicolumn{1}{c}{BAGL} & \multicolumn{1}{c}{BAGR} & \multicolumn{1}{c}{BSGR}  \\ 
\hline
\multicolumn{21}{c}{p=10} 
\\
\\
L1      & \textbf{0.04} & 0.07  & 0.07 
        & 0.55 & \textbf{0.49} & \textbf{0.49}  
        & 0.62 & 0.88 & \textbf{0.50} 
        & 0.23  & 0.30 & \textbf{0.20}   
        & 0.24 & 0.19  & \textbf{0.15} 
        & 0.71  & \textbf{0.54} & 0.55 \\

L2      & 0.05 & 0.05  & 0.05 
        & 0.54 & \textbf{0.48} & \textbf{0.48} 
        & 0.55 & 0.67 & \textbf{0.50} 
        & 0.16 & 0.22 & \textbf{0.14} 
        & 0.20 & 0.12 & \textbf{0.10} 
        & 0.66  & \textbf{0.41} & 0.46 \\ 
        
EL1     & 0.01 & 0.01 & 0.01
        & 0.05 & 0.05 & 0.05
        & 0.09 & 0.09 & \textbf{0.06} 
        & \textbf{0.03} & 0.04 & \textbf{0.03} 
        & 0.03 & \textbf{0.02} & \textbf{0.02} 
        & 0.07 & 0.06 & \textbf{0.05} \\
        
EL2     & 0.01 & 0.01 & 0.01
        & 0.16 & \textbf{0.09} & 0.11    
        & 0.26 & \textbf{0.23} & 0.33 
        & \textbf{0.02} & 0.05 & 0.03
        & 0.03 & \textbf{0.01} & \textbf{0.01} 
        & 0.30 & \textbf{0.12} & 0.16 \\
        
MAXEL1  & 0.05 & 0.05 & 0.05 
        & 0.54 & \textbf{0.49} & \textbf{0.49}  
        & 0.60 & 0.81 & \textbf{0.51} 
        & \textbf{0.15} & 0.32 & 0.24 
        & 0.25 & 0.20 & \textbf{0.17} 
        & 0.74 & 0.59 & \textbf{0.56} \\
        
MINEL1  & 0.02 & 0.02 & 0.02 
        & 0.01 & 0.01 & 0.01 
        & 0.01 & 0.01  & 0.01 
        & \textbf{0.03} & 0.04 & 0.04 
        & 0.01 & 0.01 & 0.01 
        & 0.03 & 0.03 & 0.03 \\ 
        
STEIN  & \textbf{0.01} & 0.02 & 0.02 
        & 0.03 & 0.03 & 0.03 
        & 0.03 & \textbf{0.02} & \textbf{0.02} 
        & 0.02 & 0.02 & 0.02 
        & 0.02 & 0.02 & 0.02 
        & 0.03 & \textbf{0.03} & 0.03 \\
\\
\multicolumn{21}{c}{p=30}  
\\
\\
L1      & \textbf{0.07} & 0.13  & 0.13 
        & \textbf{0.03} & 1.78 & 0.46  
        & \textbf{0.24} & 2.56 & 0.45 
        & 2.91  & 1.22 & \textbf{0.46}   
        & \textbf{0.02} & 0.79  & 0.39 
        & \textbf{0.04} & 1.84 & 0.52 \\ 
        
L2      & \textbf{0.06} & 0.07  & 0.07 
        & \textbf{0.04} & 1.67 & 0.43 
        & \textbf{0.43} & 2.13 & 0.26 
        & 2.62 & 0.84 & \textbf{0.35} 
        & \textbf{0.02} & 0.53 & 0.30 
        & \textbf{0.03} & 1.32 & 0.36 \\ 
        
EL1     & 0.01 & 0.01 & 0.01 
        & \textbf{0.01} & 0.06 & 0.02
        & \textbf{0.03} & 0.09 & 0.02 
        & 0.11 & 0.05 & \textbf{0.02} 
        & \textbf{0.01} & 0.03 & 0.02 
        & \textbf{0.01} & 0.06 & 0.02 \\
        
EL2     & 0.01 & 0.01 & 0.01
        & \textbf{0.11} & 0.38 & 0.33    
        & 1.91 & 0.74 & \textbf{0.68} 
        & 2.26 & 0.23 & \textbf{0.12}
        & 0.12 & 0.08 & \textbf{0.04} 
        & \textbf{0.10} & 0.43 & 0.34 \\
        
MAXEL1  & \textbf{0.05} & 0.06 & 0.06 
        & \textbf{0.06} & 1.75 & 0.43  
        & 0.47 & 2.59  & \textbf{0.25} 
        & 0.54 & 0.99 & \textbf{0.38} 
        & \textbf{0.19} & 0.75 & 0.38 
        & \textbf{0.05} & 1.81 & 0.36 \\
        
MINEL1  & 0.03 & \textbf{0.02} & \textbf{0.02} 
        & 0.01 & 0.01 & 0.01 
        & 0.01 & 0.01 & 0.01 
        & 0.02 & 0.02 & 0.02 
        & 0.01 & \textbf{0.01} & \textbf{0.01} 
        & 0.03 & \textbf{0.02} & \textbf{0.02} \\ 
        
STEIN  & \textbf{0.03} & 0.07 & 0.07 
        & 4.31 & 0.08 & \textbf{0.06} 
        & 1.32 & \textbf{0.07} & 0.08 
        & 3.10 & \textbf{0.06} & \textbf{0.06} 
        & 0.53 & \textbf{0.07} & \textbf{0.07} 
        & 0.55 & \textbf{0.07} & \textbf{0.07} \\
\\
\multicolumn{21}{c}{p=50}                                                      \\&&&&&&&&&&&&&&&&&&&&\\
L1      & \textbf{0.07} & 0.12  & 0.12 
        & \textbf{0.02} & 3.56 & 0.48  
        & \textbf{0.11} & 4.16 & 0.44 
        & \textbf{0.06}  & 2.23 & 0.51   
        & \textbf{0.02} & 1.76 & 0.51 
        & \textbf{0.04}  & 3.69 & 0.39 \\

L2      & \textbf{0.06} & 0.08  & 0.08 
        & \textbf{0.03} & 3.24 & 0.39 
        & \textbf{0.20} & 3.02 & 0.24 
        & \textbf{0.04} & 1.56 & 0.35 
        & \textbf{0.02} & 1.21 & 0.41 
        & \textbf{0.02} & 2.29 & 0.21 \\ 
        
EL1     & 0.01 & 0.01 & 0.01 
        & \textbf{0.01} & 0.07 & \textbf{0.01}
        & \textbf{0.01} & 0.08 & \textbf{0.01} 
        & \textbf{0.01} & 0.05 & \textbf{0.01}
        & \textbf{0.01} & 0.03 & \textbf{0.01} 
        & \textbf{0.01} & 0.06 & \textbf{0.01} \\
        
EL2     & 0.01 & 0.01 & 0.01
        & \textbf{0.13} & 0.97 & 0.38    
        & 1.95 & 1.18 & \textbf{0.78} 
        & \textbf{0.10} & 0.55 & 0.20
        & 0.19 & 0.22 & \textbf{0.10} 
        & \textbf{0.10} & 0.76 & 0.27 \\
        
MAXEL1  & \textbf{0.05} & 0.06 & 0.06 
        & \textbf{0.08} & 3.52 & 0.39  
        & \textbf{0.45} & 3.73 & 0.23 
        & \textbf{0.07} & 1.72 & 0.32 
        & \textbf{0.30} & 1.58 & 0.44 
        & \textbf{0.05} & 2.98 & 0.21 \\
        
MINEL1  & 0.03 & \textbf{0.01} & \textbf{0.01} 
        & 0.01 & 0.01 & 0.01 
        & 0.01 & 0.01 & 0.01 
        & 0.03 & \textbf{0.01} & \textbf{0.01} 
        & 0.01 & \textbf{0.01} & \textbf{0.01} 
        & 0.03 & \textbf{0.02} & \textbf{0.02} \\ 
        
STEIN  & \textbf{0.04} & 0.10 & 0.10 
        & 6.61 & 0.23 & \textbf{0.12} 
        & 3.69 & \textbf{3.24} & 3.84 
        & 0.75 & 0.11 & \textbf{0.10} 
        & 0.65 & 0.15 & \textbf{0.14} 
        & 0.77 & \textbf{0.12} & \textbf{0.12} \\

\hline
\end{tabular}
\caption{Summary of L1, L2, EL1, EL2, MAXEL1, MINEL1 and STEIN for a diagonal, AR($1$), AR($p-3$), a cluster model with two equally sized clusters, a cluster model with two clusters with a 1:5 size ratio and a full  model. The standard error of the loss values reported here are based on 30 replications for both the BAGL, BAGR and BSGR estimators. The best performing values are boldfaced.}
\label{tab:syn_loss_results_sd}
\end{sidewaystable}

\clearpage

\section{Cell signalling illustration}\label{sec:flow_cytometry}
\begin{figure}[H]

\begin{minipage}{.5\linewidth}
\centering
\subfloat[]{\label{fig:flow_cytometry_bagl}\includegraphics[scale=.082]{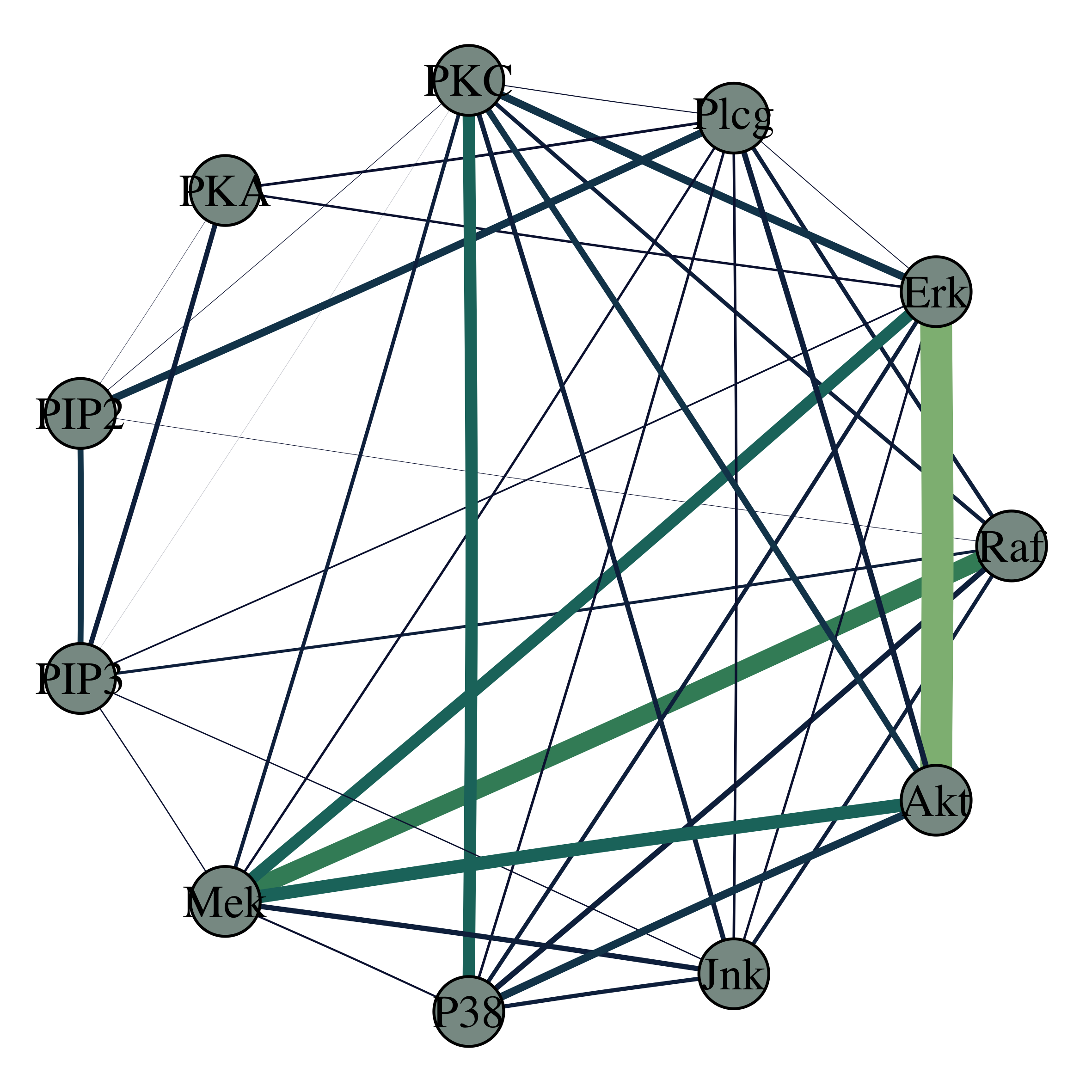}}
\end{minipage}%
\begin{minipage}{.51\linewidth}
\centering
\subfloat[]{\label{fig:flow_cytometry_bagr}\includegraphics[scale=.082]{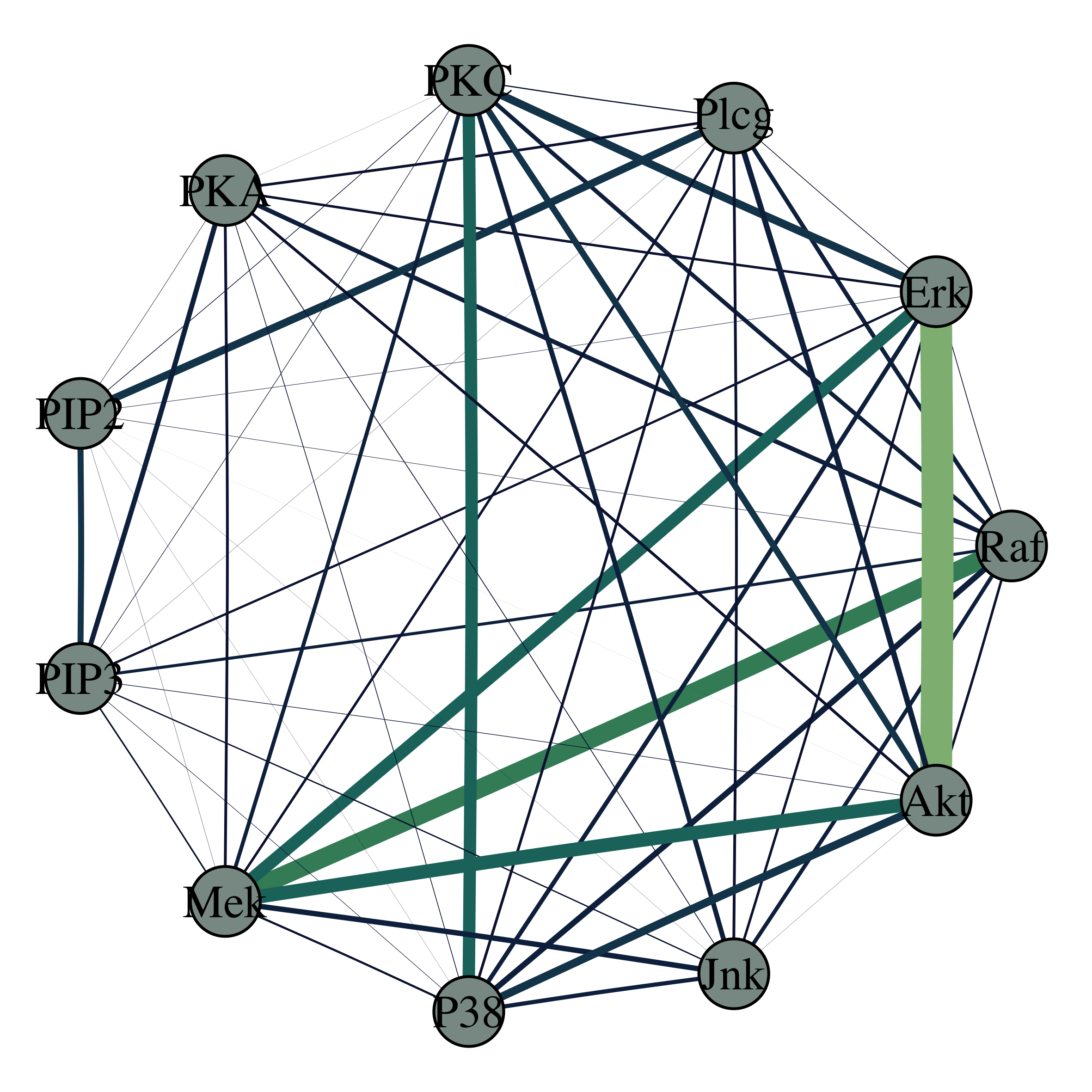}}
\end{minipage}\\[-4ex]\par\medskip
\centering
\subfloat[]{\label{fig:flow_cytometry_diff}\includegraphics[scale=.082]{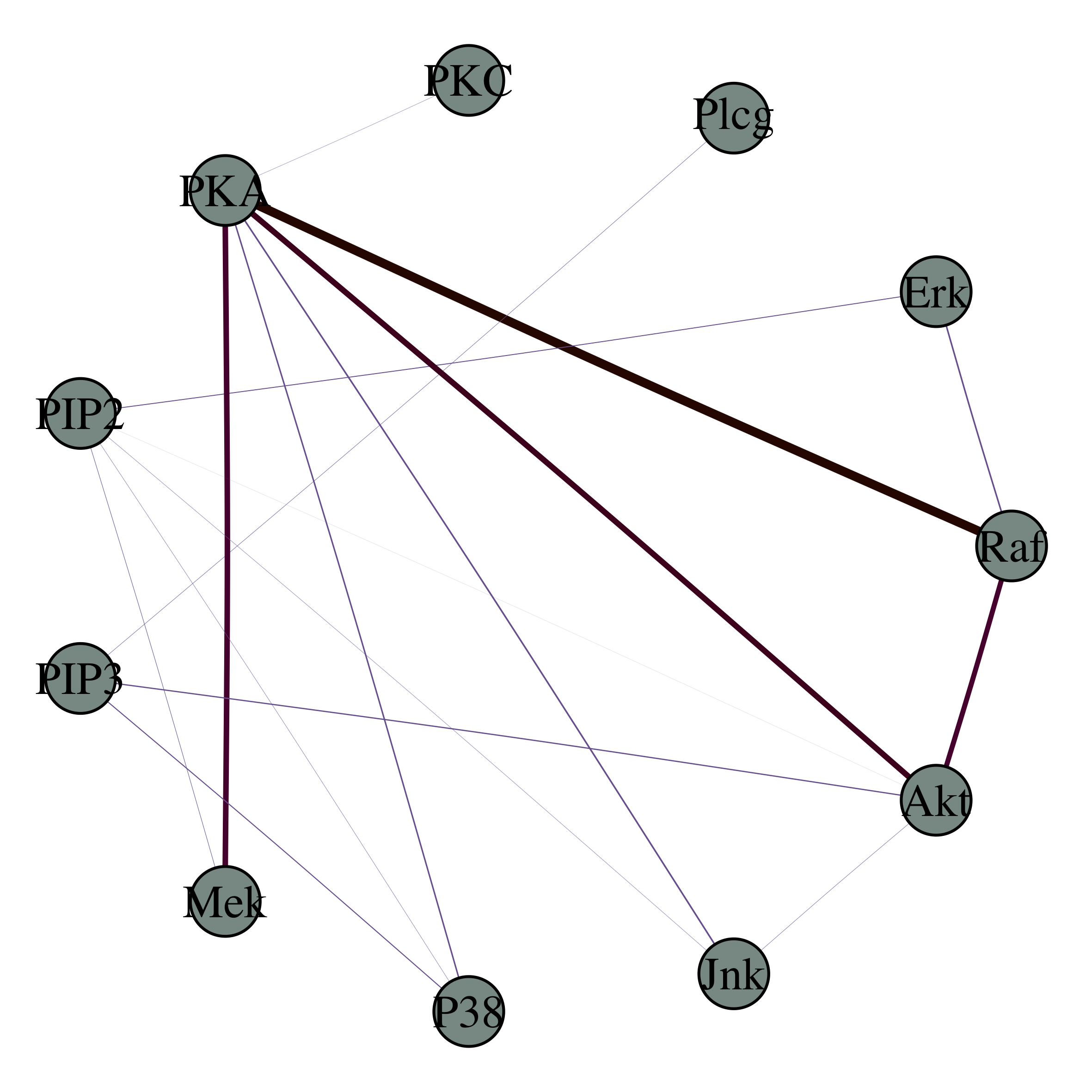}}

\caption{Undirected cell-signaling graphs estimated using the BAGL (a), BAGR (b), and the difference between the BAGL and BAGR (c).}
\label{fig:flow_cytometry_graphs}
\end{figure}

For illustration on structure learning of the BAGR and BAGL, the flow cytometry dataset from \cite{sachs2005causal} is analysed using the Bayesian adaptive graphical lasso and Bayesian adaptive graphical ridge. The goal of the modeling exercise is to provide inference on the signalling network that connects key phosphorylated proteins in human T cell signalling. The data consist of $p = 11$ proteins and $n = 7466$ cells. Flow cytometry data sets  typically present thousands of individual cells worth of independent observations making it an ideal application for ridge based modeling. The Bayesian undirected graphical models are based on the MCMC outputs consisting of 5000 iterations after 2000 burn-ins. \cite{friedman_2008_sparse} suggest using the unregularised model based on the lowest cross validation error obtained in search of the optimal shrinkage parameter. This suggestion supports the motivation and intent of the BAGR estimator. Upon inspection of Figure \ref{fig:flow_cytometry_graphs}, the BAGL estimator produces a sparser representation of the graphical model, Figure \ref{fig:flow_cytometry_bagl}, when compared to the BAGR in Figure \ref{fig:flow_cytometry_bagr}. The width of the edges represent the strength of the associations between the nodes. Figure \ref{fig:flow_cytometry_diff} highlights the associations that the BAGR is capable of bringing to light, where the BAGL cannot. Noticeably, the BAGL estimator struggles to map a few well known connections such as the direct enzyme-substrate relationships between PKA and Raf, as well as the phosphorylation association represented by the relationships between Plcg and PIP2. The BAGR estimator, similarly to the Bayesian network used in \cite{sachs2005causal}, is also capable of detecting the indirect connections represented by the relationships between PKA and P38, as well as between PKA and Jnk. Lastly, the BAGR model captures a connection that is not mapped by either the BAGL nor the Bayesian network, namely the relationships between PIP3 and Akt.

\section{Discussion}\label{sec:discussion}
The Bayesian graphical ridge-type estimators are shown to be attractive for estimating Gaussian graphical models when prioritisation is placed on accurate association representation over sparsity and reducing the computational footprint with an increasing feature space. The Bayesian graphical ridge-type estimators also offer insight and inference into the estimation procedure, via the MCMC results, when compared to frequentist counterparts. The Bayesian graphical ridge-type estimators also enjoy low computational costs for low to moderate dimensions due to the efficient block Gibbs sampler, adding to the Bayesian toolbox of precision matrix estimation. Synthetic studies indicate strong empirical evidence, in favour of the Bayesian graphical ridge-type estimators, for the estimation of precision matrices that are relatively non-sparse. The Bayesian adaptive graphical ridge-type estimator demonstrates the ability to successfully infer the associations between key phosphorylated proteins in human T cell signalling that may provide valuable clinical inference, for example in understanding responses to complex drug therapies used to treat cancer. Finally, with regards to graphical structure learning, the Bayesian graphical ridge-type estimators cannot perform graphical structure determination, an obvious requirement for larger dimensions. As a result, a Bayesian elastic net prior is currently being developed to address the latter whilst maintaining the association accuracy of the Bayesian graphical ridge-type estimators, as well as its computational efficiency.    

\section*{Funding}
This work was based upon research supported in part by the National Research Foundation (NRF) of South Africa, SARChI Research Chair UID: 71199; Ref.: IFR170227223754 grant No. 109214; Ref.: SRUG190308422768 grant No. 120839. The opinions expressed and conclusions arrived at are those of the authors and are not necessarily to be attributed to the CoE-MaSS or the NRF. The research of the corresponding author is supported by a grant from Ferdowsi University of Mashhad (N.2/58091).



\bibliographystyle{tfs}
\bibliography{interacttfssample}

\end{document}